\documentclass[sigconf]{acmart}
\usepackage{algorithm}
\usepackage[noend]{algpseudocode}
\AtBeginDocument{%
  }

\setcopyright{acmlicensed}
\copyrightyear{2026}
\acmYear{2026}
\acmConference[MIG '26]{ACM SIGGRAPH Conference on Motion, Interaction and Games}{December 11--13,
  2026}{Charleston, SC}
\acmISBN{978-1-4503-XXXX-X/2026/12}

\begin{document}

\title{
Endless VR Walk: Dynamic Mapping of Limited Physical Spaces into Infinite Linear Paths
}

\author{Sofia Martinez}
\affiliation{%
  \department{Digital Worlds Institute}
  \institution{University of Florida}
  \city{Gainesville}
  \state{Florida}
  \country{USA}
}

\author{Reagan Haddix}
\affiliation{%
  \department{Digital Worlds Institute}
  \institution{University of Florida}
  \city{Gainesville}
  \state{Florida}
  \country{USA}
}

\author{Miles Haase}
\affiliation{%
  \department{Digital Worlds Institute}
  \institution{University of Florida}
  \city{Gainesville}
  \state{Florida}
  \country{USA}
}

\author{Angelos Barmpoutis}
\affiliation{%
  \department{Digital Worlds Institute}
  \institution{University of Florida}
  \city{Gainesville}
  \state{Florida}
  \country{USA}
}

\renewcommand{\shortauthors}{Martinez et al.}

\begin{abstract}
 Natural locomotion in virtual reality (VR) can provide a more immersive and engaging experience, but the physical space required for unrestricted walking is a significant limitation. This paper presents Endless VR Walk, a system for dynamically mapping a limited physical space into an effectively infinite linear virtual path. The approach uses a recursive path-generation algorithm to construct virtual blocks in real time as the user walks, while constraining the path to the available physical area and preventing spatial overlap. We evaluate path variability under different physical-space constraints and identify the smallest space that enables diverse path configurations. The resulting framework provides a platform for applications benefiting from extended walking, including physical therapy, rehabilitation, and fitness, as well as adaptations of traditional platformer and endless-runner games in VR. 
\end{abstract}

\ccsdesc[500]{Software and its engineering~Interactive games}
\ccsdesc[500]{Human-centered computing~Virtual reality}

\keywords{Virtual Reality, Locomotion, Infinite Walk, Procedural Generation, Games, Impossible Spaces}
\begin{teaserfigure}
 \includegraphics[width=\textwidth]{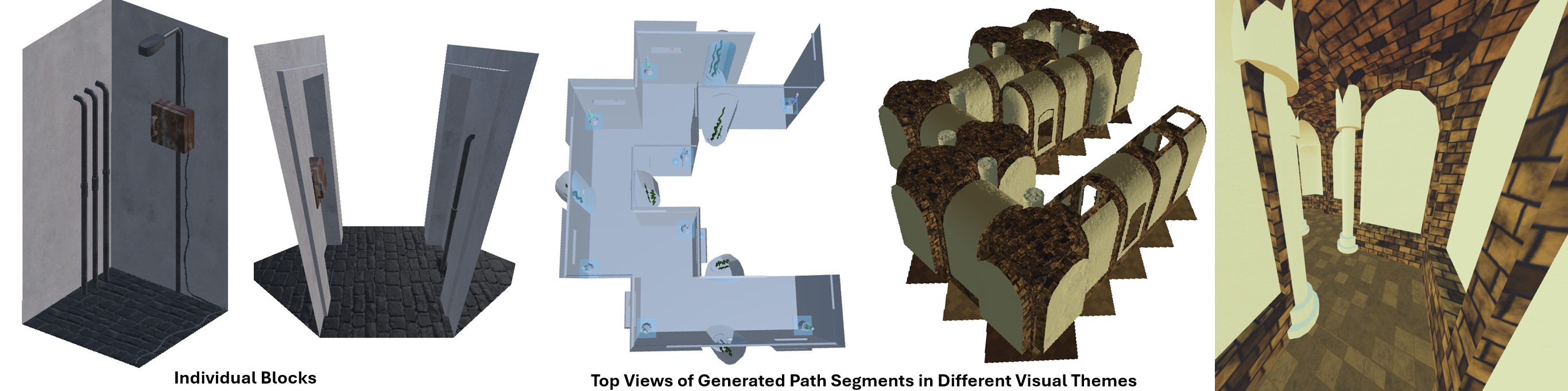}
  \caption{Screenshots of the generated path using our proposed real-time dynamic path mapping algorithm.}
  \Description{Enjoying the baseball game from the third-base
  seats. Ichiro Suzuki preparing to bat.}
  \label{fig:teaser}
\end{teaserfigure}


\maketitle

\section{Introduction}
Numerous approaches have been proposed to overcome the locomotion challenges and physical space constraints in VR. Omni-directional treadmills and artificial locomotion can help with immersion but are inaccessible to most VR users or can increase cybersickness \cite{hooks2020evaluating,walking_pad_2018,Longshortterm}. Other solutions like discrete rotation and lateral movements \cite{Farmani}, stuttered locomotion \cite{Freiwald}, and gaze predictions \cite{DavidJohn} offer alternative paths that could optimize user movement and comfort. Since freedom of interaction is an important factor in mitigating motion sickness \cite{Hao}, natural walking can provide a more comfortable and immersive alternative to artificial locomotion techniques in VR. Because physical space limits the ability to support unrestricted walking in an infinite virtual environment, recent research has explored space reuse techniques that map topologically impossible virtual spaces onto a shared physical space \cite{LochnerGain}. 

In this paper, we extend the concept of impossible spaces by proposing a method for uninterrupted locomotion along an effectively infinite linear virtual path that is dynamically generated within the boundaries of a physical environment. By relying on natural walking as the primary locomotion mechanism, our approach has the potential to enhance the sense of presence while mitigating cybersickness \cite{Hao}. Potential applications include VR video games, such as platformers and endless-runners with procedural level generation, and home-based exercise, where the illusion of unlimited walking distance could support more immersive and physically engaging experiences.

\section{Methods}
Given any space, one can perform a random walk by taking a series of steps in different directions within the space. To design a structurally sound and meaningful 3D environment along this path, design testing showed that restricting the path generation to three possible directions—forward, $90^\circ$ left turn, and $90^\circ$ right turn—made more effective use of the available space than a random walk where any direction was possible. Restricting the path generation to only three directions made it easier to design environment sets. At minimum, each set needs three building blocks, each responsible for one direction (see Fig. \ref{fig:teaser} left for examples of blocks). To populate the random path, we developed three themed asset libraries. 

In order to include objects outside the scale of the hall (alcoves, niches, etc), sharp “U” turns were not permitted. The themes chosen to test the random path were termed: Alley Way, Frutiger Aero, Castle (Fig. \ref{fig:teaser}). Through design testing, we established an appropriate scale for the block assets in the virtual environment ($1.2m \times 1.2m$), allowing us to create a sufficiently wide path while minimizing the physical space required to support the experience. To avoid users perceiving the generated paths as repetitive, we evaluated path variability as a function of the available physical space.

\begin{algorithm}
\caption{Recursive Forward Block Generation}
\label{alg:forward-generation}
\begin{algorithmic}[1]
\Function{GenerateForward}{$origin, remaining$}
    \If{$remaining = 0$}
        \State \Return success
    \EndIf

    \State $q \gets$ GetNextQuad($origin$)
    \If{not WithinBounds($q$) or OverlapsBlocks($q$)}
        \State \Return failure
    \EndIf

    \ForAll{$type$ in Shuffled(AllowedTypesAfter($origin$))}
        \State $newBlock \gets$ CreateBlock($origin, type$)
        \If{\Call{GenerateForward}{$newBlock, remaining - 1$}}
            \State \Return success
        \EndIf
        \State Remove($newBlock$)
    \EndFor

    \State \Return failure
\EndFunction
\end{algorithmic}
\end{algorithm}

We designed a recursive algorithm to dynamically generate a random path as the user walks, subject to the constraints that newly generated blocks remain within the bounds of our lab's available physical space ($6m \times 6m$) and do not overlap with previously generated blocks within a specified window ($w=5$). The user's position is continuously tracked by the head-mounted display (HMD), and when the user exits the current block, the algorithm is called as $GenerateForward(getBlock(current+5),4)$. This call uses the block five positions ahead as the origin and recursively generates four additional blocks, ensuring that five blocks remain visible ahead of the user while four additional blocks are generated as a staging area. Similarly, five previously generated blocks behind the user remain active and visible, while earlier blocks are deactivated to prevent visual overlap and reduce unnecessary rendering. This setup allows users to move freely in either direction. The structure of previously generated parts of the path are retained in memory, to be reconstructed when revisited and thereby creating the perception of a persistent virtual space despite its dynamic generation.

\section{Discussion of Preliminary Testing}
The proposed system was implemented in the Unity game engine using assets created in Maya and Blender and tested on Meta Quest 3 headsets. The source code is available at the following address: \url{https://github.com/digitalworlds/MiDAS-Infinite-Walk}.

Table \ref{tab:variability} reports the average frequencies of left, right, and straight blocks, as well as the frequency of direction changes (e.g., “S” patterns in the path) and zigzags (i.e., small alternating turns that do not produce a sustained change in the path's overall angular direction), calculated from 100 randomly generated paths within specified area limits. In the smallest scenario, a circular path is generated within a $3\times3$ block grid, resembling the setup described in \cite{tian}, but without any structural variability observed. As the available space increases, the observed path variations also increase. The smallest space in which paths exhibiting the full range of variations can be generated is a $5\times5$ block grid. 

During initial implementation, we observed noticeable lag when generating new blocks by instantiating meshes from the asset library at runtime. To reduce this overhead, we adopted an object-reuse strategy in which ten instances of each block type were pre-instantiated during initialization. When a new block was required, an existing instance was repositioned and activated rather than creating a new object. This approach reduced the computational overhead, resulting in a smoother user experience.

\begin{table}
  \caption{Random Path Variability Per Available Area}
  \label{tab:variability}
  \begin{tabular}{ccccccc}
    \toprule
    Blocks&Area&$\Lsh$&$\Rsh$&$\uparrow$&Dir.Ch.&ZigZag\\
    \midrule
    $3 \times 3$ & $12m^2$ & 50.0\% & 0.0\% & 50.0\% & 0.0\% & 0.0\% \\
    $4 \times 4$ & $23m^2$ & 43.4\% & 5.2\% & 51.3\% & 0.0\% & 10.4\%\\
    $5 \times 5$ & $36m^2$ & 25.2\% & 24.9\% & 50.0\% & 1.6\% & 20.3\%\\
    $9 \times 9$ & $117m^2$ & 26.6\% & 26.4\% & 47.0\% & 5.0\% & 32.0\%\\
  \bottomrule
\end{tabular}
\end{table}

\section{Future Directions}
Moving forward, we plan to use the system to investigate how perception of an effectively infinite virtual space influences a user's perception of the underlying physical environment. Future experiments will examine how this perception is affected by factors, including variations in environmental themes, designs, and landmarks as well as spatial variables, such as changes in corridor width, and branching pathways creating maze-like configurations. As a further extension toward a finished system, the path-generation area could be dynamically constrained to the virtual boundary defined by the HMD, allowing users to define the physical area in which the system generates the virtual path. Finally, the proposed system could be evaluated in applications involving walking, including physical therapy, rehabilitation, and fitness. 

\newpage
\bibliographystyle{ACM-Reference-Format}
\bibliography{bibliography}

\end{document}